\documentclass[a4paper,twocolumn,10pt, notitlepage]{quantumarticle}
\pdfoutput=1
\usepackage[english]{babel}
\usepackage[T1]{fontenc}

\usepackage{graphicx}% Include figure files
\usepackage{dcolumn}% Align table columns on decimal point
\usepackage{bm}% Bold math

\usepackage{physics}
\usepackage{amsmath, mathcomp, textcomp, mathrsfs, amssymb, mathtools, amsfonts, dsfont}
\usepackage{xspace, arydshln}
\usepackage{float}
\usepackage{hyperref}
\usepackage{xparse}
\usepackage{cancel}
\usepackage{pifont}

\usepackage[numbers,sort&compress]{natbib}

\newcommand{\Sx}{\ensuremath{\hat{S}_x}\xspace}
\newcommand{\Sy}{\ensuremath{\hat{S}_y}\xspace}
\newcommand{\Sz}{\ensuremath{\hat{S}_z}\xspace}

\newcommand{\Axx}{\ensuremath{A_{xx}}\xspace}
\newcommand{\Ayy}{\ensuremath{A_{yy}}\xspace}
\newcommand{\Azz}{\ensuremath{A_{zz}}\xspace}

\newcommand{\Ham}{\ensuremath{\hat{H}}\xspace}

\usepackage[table]{xcolor}
\newcommand{\xmark}{\ding{55}}

\definecolor{light_blue}{RGB}{225, 242, 250}
\definecolor{light_orange}{RGB}{255, 228, 214}

\begin{document}

\title{Entanglement Generation on the Double Quantum Transition of NV Ground State Via Globally Addressing Microwave Pulse}

\author{Marcel Morillas-Rozas}
\orcid{0009-0005-4570-1016}
\author{Alberto L\'opez-Garc\'ia}
\orcid{0009-0001-0023-9850}
\author{Javier Cerrillo}
\email{javier.cerrillo@upct.es}
\orcid{0000-0001-8372-9953}
\affiliation{\'Area de F\'isica Aplicada, Universidad Polit\'ecnica de Cartagena, Cartagena 30202, Spain}

\begin{abstract}
  Entanglement is a key quantum feature that enables quantum sensors to improve their sensitivity up to the Heisenberg limit. In the NV-center platform, the Heisenberg limit can only be achieved when the axes of the NV centers are parallel. Nevertheless, parallel NV centers are spectrally indistinguishable and no mechanisms to directly prepare Heisenberg-limit-grade entanglement in such configurations are known to date. In this work we propose for the first time a viable mechanism to prepare entangled states in the double quantum transition of two dipolarly coupled NV centers whose axes are parallel without populating intermediate states, so as to reach the Heisenberg limit in sensing. Our approach is based on the NV effective Raman coupling (NV-ERC) protocol and makes use of global addressing of both NV centers with a single monochromatic microwave pulse. Supported by an adiabatic elimination analysis, several mechanisms for the preparation of different entangled states are identified, all of which avoid the involvement of intermediate states. This not only minimizes the impact of additional noise sources, but also enables the state generation process itself to serve as effective sensing time—an advantage over conventional approaches where such preparation typically constitutes a separate, non-contributory stage. We consider the generation of different entangled states belonging to the double quantum transition, sensitive to either transverse electric fields or longitudinal magnetic fields, all with a fourfold improved sensitivity compared to conventional single NV settings.
\end{abstract}

\maketitle

The field of quantum sensing using negatively charged nitrogen-vacancy (NV) centers in diamond \cite{NV_Centers_Fedor, PhysRevLett.92.076401, DOHERTY20131} has seen a rapid growth due to the ability of NV centers to perform high-precision measurements of magnetic \cite{Magnetic_Glenn, Magnetic_Cappellaro, Magnetic_Webb, Magnetic_Zheng, Magnetic_Muller}, electric \cite{Electric_Dolde, Electric_Chen} and thermal fields \cite{Temperature_Sensing1, Temperature_Hui} with nanoscale precision \cite{PhysRevLett.112.097603, HU2024111858, PhysRevB.107.L140101}. Moreover, the ground state of the NV center shows long coherence times of the spin-1 triplet ground state \cite{PhysRevB.101.184430, PhysRevApplied.20.044045, PhysRevB.105.205401}, even at ambient temperatures, and they may be optically initialized and read out. Measurements can be performed extraordinarily close to the sample \cite{QuantumSensing_Cappellaro}, especially in the case of nanodiamonds. These features position NV centers above some of their competitors, due to the fact that experimental NV center implementations do not require vacuum chambers or complex laser arrangements, and can be operated at room temperature.\\

Although the NV center ground state is a spin-1 triplet, the conventional approach is to treat it as a 2 level system by spectral selection of one of the transitions \cite{MAUDSLEY1986488, PhysRevLett.106.240501, PhysRevA.83.022306, kotler2011single, PhysRevB.77.174509, PhysRevLett.98.100504, doi:10.1126/science.1192739, PhysRevLett.105.200402, PhysRevB.83.081201, PhysRevA.92.042304, abobeih2018one, PhysRevLett.122.200403, Fedor_Entanglement, Dolde2013}. However, this approach requires the Zeeman splitting of the double quantum transition to be larger than the Rabi frequency of the control microwave (MW) pulse. If this condition is not satisfied, population leakage to the third state results in sensing limitations. Specifically, high frequency sensing and low bias magnetic field operation become unfeasible. These limitations are resolved by the NV effective Raman coupling (NV-ERC) approach \cite{PhysRevLett.126.220402, PhysRevApplied.17.044028, PhysRevApplied.21.054011, Paper_Alberto}, which allows to design operations that account for all three levels of the NV center without any loss of population.\\

One way to enhance sensing protocols is to exploit quantum entanglement, which allows to surpass the precision imposed by the standard quantum limit (SQL) and pushes it towards the Heisenberg limit. NV sensing protocols achieving a precision above the SQL have already been demonstrated \cite{Science_Sub_SQL, liu2015demonstration}. In particular, entanglement between a single NV center's electronic and nuclear spins have been demonstrated \cite{Entanglement_NV_nucleus, PhysRevLett.107.150503}, as well as entanglement between two distant NV centers's electronic spins mediated by dipolar interactions and addressing the transition of each NV center individually \cite{Fedor_Entanglement} by means of spectral selection. This would not be possible in the low field limit and requires a pair of NV centers that are aligned in different axes. However, the Heisenberg limit cannot be reached unless both NV centers couple identically to the external perturbation, which requires their axes to be parallel. Misalignment leads to non-identical Zeeman splittings, resulting in different phase accumulation in each NV center. In this case, the addition of the phases of the $N$ spins does not scale with $N$, which is a fundamental requisite to reach the Heisenberg limit. For the NV centers to couple identically to the perturbation and thus enable reaching the Heisenberg limit, their axes must be parallel. In this configuration, however, they become spectrally indistinguishable, precluding the use of conventional spectral selection techniques.\\

Theoretical proposals for the generation of entanglement between two or more parallel NV centers exist that do not reach the Heisenberg limit \cite{hybrid_1, hybrid_2} or require a symmetry that abruptly breaks down in the presence of magnetic fields \cite{Annealing_Matsuzaki}, rendering it unsuitable for sensing.\\

Using the NV-ERC approach, we propose a protocol to overcome this limitation which allows us to generate several maximally entangled states that belong to the double quantum transition of the NV center's ground state in a system comprising two parallel NV centers interacting via dipole interaction. The system is considered to be under the effect of an external magnetic field and is globally addressed with a single monochromatic MW field. By means of a Raman transfer, the entangled state is generated directly from the ground state, with almost negligible participation of intermediate states. Avoiding intermediate states is crucial to suppress unwanted perturbations and reaching the Heisenberg limit. Therefore, Raman transfer is a well justified approach from a quantum sensing perspective since unwanted phase collection from intermediate states is avoided. All possible geometrical configurations of the two NV-centers are analyzed. An alternative approach in the limit of vanishing bias magnetic field is explored with similar efficiency. This proposal enables for the first time the creation of Heisenberg-limited sensing states for NV centers.\\

This manuscript is organized as follows: In Sec.~\ref{sec:preliminaries} we set out the two-NV-center model and establish a suitable picture for its analysis. Then, in Sec.~\ref{sec:results} we present three protocols for the generation of entangled states of the double quantum transition of the two NV centers. The relevant adiabatic elimination procedure is sketched in all cases, together with an analysis of the robustness of the protocol as a function of the relevant parameters. Lastly, in Sec.~\ref{sec:conclusions} we present some concluding remarks.

\section{\label{sec:preliminaries}Model}
\begin{figure}[h]
    \centering
    \includegraphics[trim=14cm 0cm 14.5cm 0cm, clip,width=\linewidth]{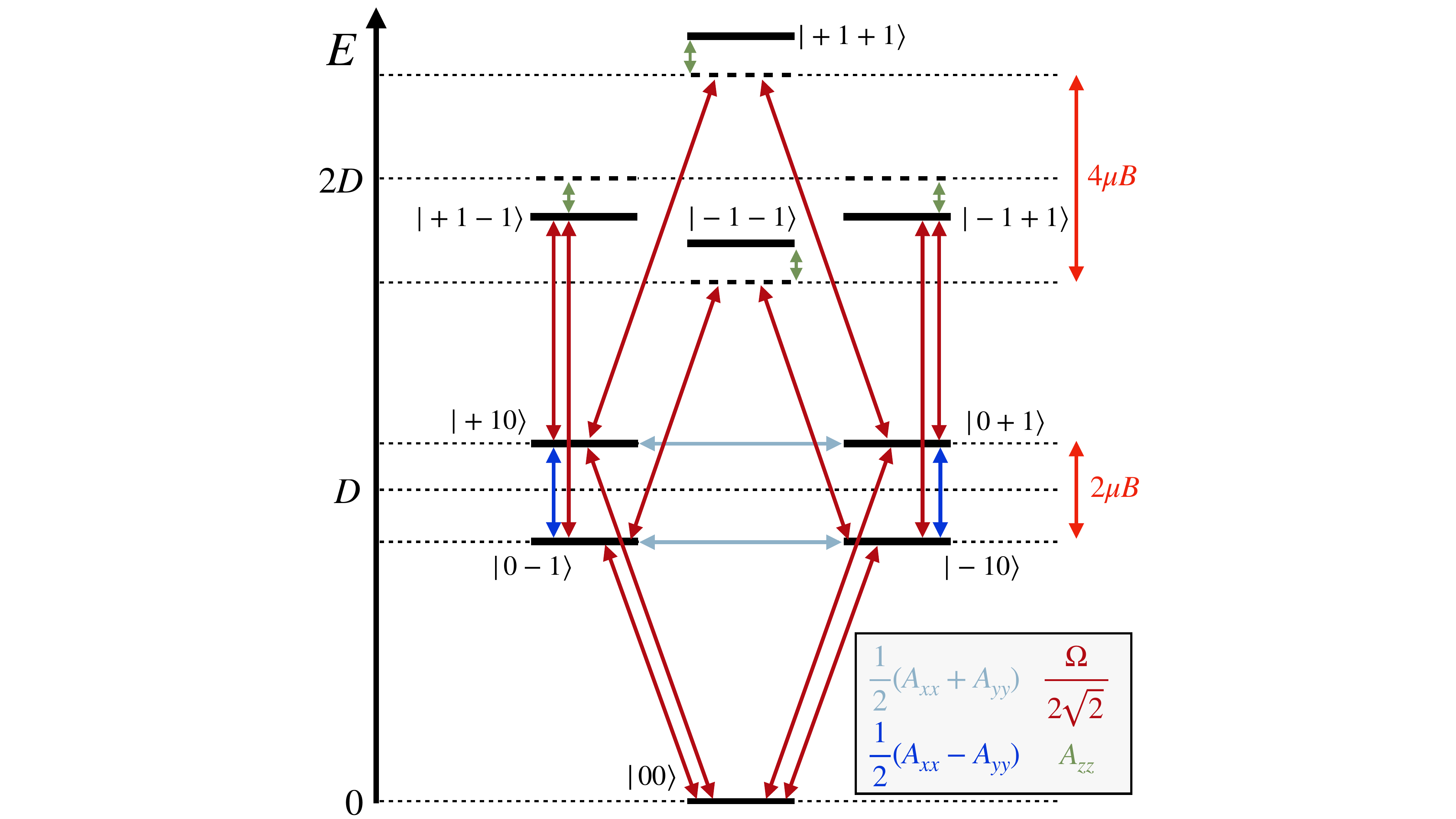}
    \caption{Energy levels of the 2 NV center system in the canonical basis after moving to a rotating frame and performing the RWA. The wine color arrows represent the couplings due to the MW field and have a value of $\Omega/2\sqrt{2}$. The olive green arrows are shifts caused by the dipolar interactions $\Sz\otimes\Sz$ and have a value of $\Azz$. The gray and the blue arrows correspond to the couplings caused by the dipole-dipole interaction $\Sx\otimes\Sx$ and $\Sy\otimes\Sy$. They have values of $\frac{1}{2}(\Axx+\Ayy)$ (gray couplings) and $\frac{1}{2}(\Axx-\Ayy)$ (blue couplings). $\Azz$ has been assumed to be positive in this figure. For the sake of simplicity, $\Delta = \omega-D$ has been set to zero in this figure.}
    \label{fig:energy_original}
\end{figure}

\begin{figure}[H]
    \centering
    \includegraphics[width = 0.92\linewidth]{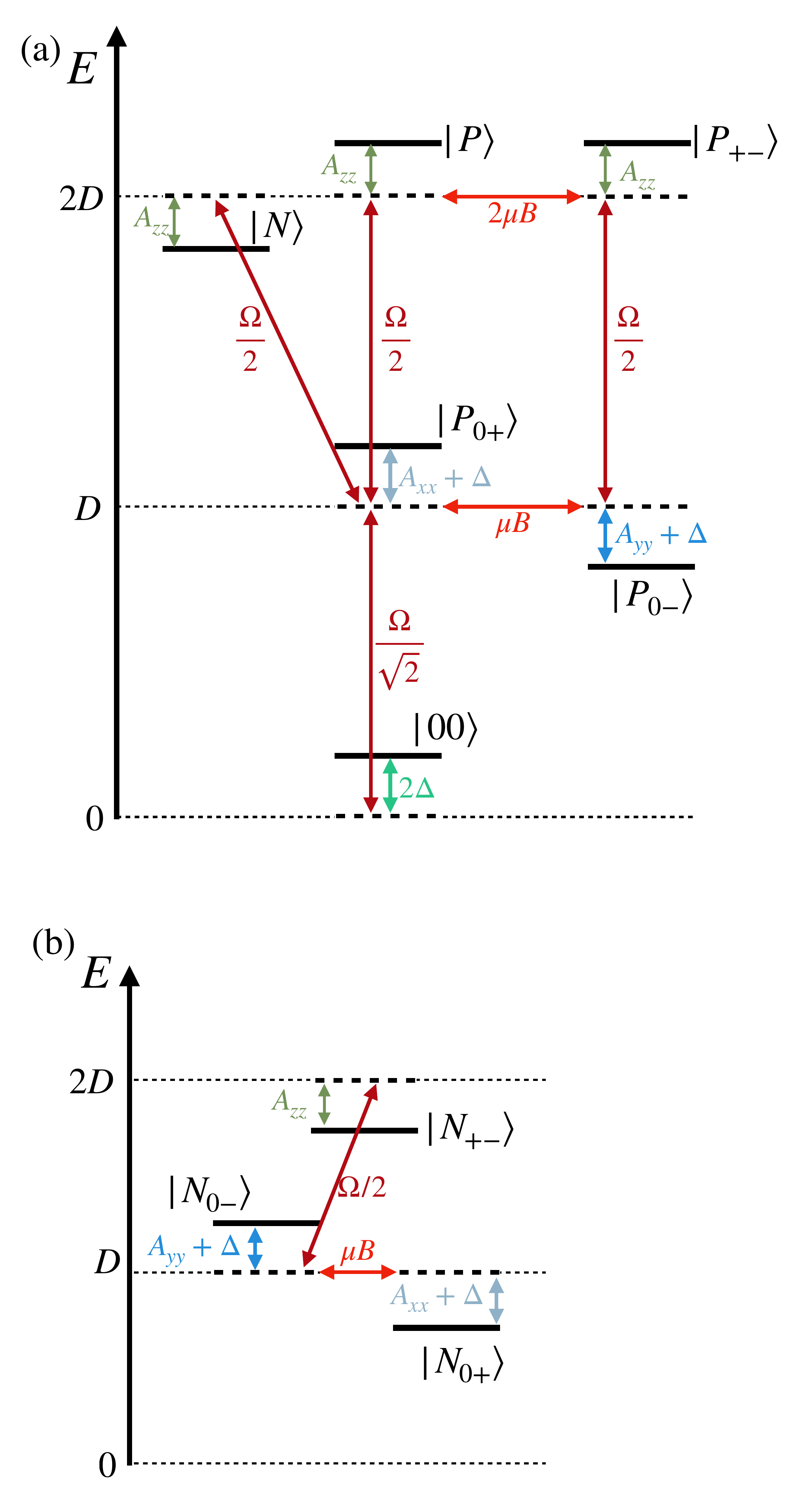}
    \caption{Energy levels of the system of two interacting NV centers under the effect of an external magnetic field and driven by a MW pulse in the new basis. (a) contains the six states that can be reached from the ground state $\ket{00}$ (bright subspace) and (b) contains the three states that become isolated (dark subspace). In red the couplings caused by the external magnetic field, in wine color the couplings coming from the driving MW field, in olive green the shifts caused by the $\Sz\otimes\Sz$ dipole-dipole interaction, in gray the energy shifts caused by the dipole-dipole interaction $\Sx\otimes\Sx$ and the $\Delta$ detuning, in blue the energy shifts caused by the $\Sy\otimes\Sy$ interaction and the detuning and in turquoise the shift caused by the detuning $\Delta$ in the ground state.}
    \label{fig:energy_diagram}
\end{figure}

The system under study consists of two NV centers $i=1,2$ whose axes are parallel, making them spectrally indistinguishable. Their ground state is a spin $S_i=1$ triplet subject to the effect of a magnetic field $\mathbf{B} = (0,0,B)^T$. The Hamiltonian describing the $i$-th NV center in angular units ($\hbar = 1$) reads
\begin{equation}
    \Ham_{\text{NV}_i} = D\hat{S}_{i,z}^2 + \mu \mathbf{B}\cdot\mathbf{S}_i = D\hat{S}_{i,z}^2 + \mu B\hat{S}_{i,z},
\end{equation}
where $D/(2\pi) = 2.87$ GHz is the zero-field splitting of the electronic ground state, $\mu /(2\pi)\approx 28$ GHz/T is the gyromagnetic factor and $\mathbf{S}_i = (\hat{S}_{i,x}, \hat{S}_{i,y}, \hat{S}_{i,z})$, where $\hat{S}_{i,j}$ are the spin-1 matrices in the $j$-th axis of NV center $i$. The two NV centers are coupled via dipole-dipole interaction, whose Hamiltonian reads
\begin{equation}\label{eq:H_dip}
    \Ham_{\text{dip}} = \frac{\mu_0\mu^2}{4\pi r^3}\left[\mathbf{S}_1\cdot\mathbf{S}_2 - 3(\mathbf{S}_1\cdot\hat{\mathbf{r}})(\mathbf{S}_2\cdot\hat{\mathbf{r}})\right],
\end{equation}
where $\mu_0$ is the vacuum permeability and $\hat{\mathbf{r}} = (\sin\theta\cos\phi, \sin\theta\sin\phi, \cos\theta)^T$ is the unit vector that points from the first NV center to the second NV center in spherical coordinates. Without any loss of generality, we consider the second NV center to be contained in the $x-z$ plane. Therefore, the unit vector in Eq.~\eqref{eq:H_dip} becomes $\mathbf{\hat{r}} = (\sin\theta, 0, \cos\theta)^T$ in spherical coordinates, where $\theta$ is the angle with respect to $\hat{\mathbf{e}}_z$. We can expand the Hamiltonian in Eq.~\eqref{eq:H_dip} as $\Ham_{\text{dip}} = \sum_{j,k}A_{jk}\hat{S}_j\otimes\hat{S}_k$, with $j,k = x,y,z$ and
\begin{equation}
    A_{jk} = \frac{\mu_0\mu^2}{4\pi r^3}\left[\delta_{jk} - 3\hat{\mathbf{r}}_j\hat{\mathbf{r}}_k\right],
\end{equation}
where $\delta_{jk}$ is Kronecker's delta. A direct consequence of our choice of having both NV centers in the $x-z$ plane is that all the $A_{jk}$ coefficients that involve the coordinate $y$ vanish except for $\Ayy$. Moreover, the cross terms $A_{jk}\hat{S}_j\otimes\hat{S}_k$ with $j\neq k$ involving $x$ and $z$ are different from zero but may be neglected under the rotating wave approximation (RWA) after transforming our Hamiltonian into an appropriate rotating frame. Therefore, the only surviving terms in the dipole-dipole interaction Hamiltonian after applying the RWA are those of the form $A_{jj}\hat{S}_{j}\otimes\hat{S}_{j}$, with $j = x,y,z$ (see Appendix~\ref{sec:appendix_dipole} for a detailed derivation). $A_{jj}$ coefficients take the form $A_{jj} = \frac{\mu_0\mu^2}{4\pi r^3}\left(1-3\hat{\mathbf{r}}^2_j\right)$. It is worth noticing that $\Axx + \Azz = -\Ayy$ in our setting. The two NV centers are driven by a MW pulse of amplitude $\Omega$ and frequency $\omega$, as shown in the Hamiltonian
\begin{equation}
    \Ham_{\text{MW}} = \Omega\cos(\omega t)\left[\hat{S}_{1,x} + \hat{S}_{2,x}\right].
\end{equation}
Hence the full Hamiltonian of the system reads
\begin{equation}\label{eq:full_H}
    \Ham = \Ham_{\text{NV}_1} + \Ham_{\text{NV}_2} + \Ham_{\text{dip}} + \Ham_{\text{MW}}.
\end{equation}
To resolve its time dependence, we move to a rotating frame relative to $\Ham_0 = \omega\left(\hat{S}_{1,z}^2 + \hat{S}_{2,z}^2\right)$ and apply the RWA, resulting in 
\begin{align}
    \Ham =& \mu B(\hat{S}_{1,z}+\hat{S}_{2,z}) - \Delta(\hat{S}^2_{1,z}+\hat{S}^2_{2,z}) + \frac{\Omega}{2}(\hat{S}_{1,x} + \hat{S}_{2,x})\notag\\
    +&\left(\Axx\ketbra{0+}{+0} + \Ayy\ketbra{0-}{-0} + \text{H.c}\right)\notag\\
    \label{eq:H_RWA}
    +&\Azz\hat{S}_{1,z}\hat{S}_{2,z},
\end{align}
where $\ket{\pm} = \frac{1}{\sqrt{2}}(\ket{1} \pm \ket{-1})$ and $\Delta = \omega - D$ have been defined. The couplings are represented graphically in Fig.~\ref{fig:energy_original} in the canonical basis, where  $\Delta = 0$ is assumed for simplicity. The complexity of the system is apparent, and the choice of a suitable basis is essential to inform the ideation of control protocols.
We propose the change to the following basis
\begin{align}
    \label{eq:B++}
   \ket{P} = & \frac{1}{\sqrt{2}}(\ket{++} + \ket{--}),\\
    \label{eq:D++}
    \ket{N} =& \frac{1}{\sqrt{2}}(\ket{++} - \ket{--}),\\
    \ket{P_{ij}} = &\frac{1}{\sqrt{2}}(\ket{ij} + \ket{ji}),\hspace{.5cm} i\neq j\\
    \ket{N_{ij}} = &\frac{1}{\sqrt{2}}(\ket{ij} - \ket{ji}),\hspace{.5cm} i\neq j
\end{align}
where $i,j\in\lbrace0, \pm\rbrace$, and state $\ket{00}$ completes the basis. Note that in the canonical basis, the $\ket{P}$ and $\ket{P_{+-}}$ states correspond to GHZ states \cite{GHZ} in the double quantum transition. The couplings in this new basis are graphically depicted in Fig.~\ref{fig:energy_diagram}. The Hamiltonian in the new basis contains a subspace of three states ($\ket{N_{+-}}, \ket{N_{0-}}$ and $\ket{N_{0+}}$) which are completely decoupled from the rest of states, i.e. it is orthogonal to the rest of the system. Therefore, unless our initial state is in this subspace, these states cannot be reached. We can split the Hamiltonian from Eq.~\eqref{eq:H_RWA} in a term for the decoupled (d) or dark subspace and one for the rest of the system, or bright subspace (b), so
\begin{equation}
    \Ham' = \Ham_\text{b} + \Ham_\text{d}.
\end{equation}
In turn, the bright subspace Hamiltonian can be expressed as the sum of three terms
\begin{equation}\label{eq:H_RWA_DARK}
    \Ham_{\text{b}} = \Ham_{\text{shift}} + \Ham_{\text{Z}} + \Ham_{\text{MW}},
\end{equation}
containing the coupling and detuning shifts
\begin{align}
    \Ham_{\text{shift}} =& (A_{xx} + \Delta)\ketbra{P_{0+}} + (A_{yy} + \Delta)\ketbra{P_{0-}}\notag\\
    +&A_{zz}(\ketbra{P} - \ketbra{N} + \ketbra{P_{+-}})\notag\\
    +&2\Delta\ketbra{00},
 \end{align}
the Zeeman splitting
  \begin{align}
    \Ham_{\text{Z}} = &\mu B\left(2\ketbra{P}{P_{+-}} + \ketbra{P_{0+}}{P_{0-}}+\text{H.c}\right),
\end{align}
and a third term for the MW driving field
\begin{align}
    \Ham_{\text{MW}} =&\frac{\Omega}{2}\Big[\sqrt{2}\ketbra{00}{P_{0+}} + \ketbra{N}{P_{0+}} + \ketbra{P}{P_{0+}}\notag\\
    &+\ketbra{P_{+-}}{P_{0-}} + \text{H.c}\Big].
\end{align}
The dark subspace Hamiltonian reads
\begin{align}
    \Ham_{\text{d}} = &\left[\mu B\ketbra{N_{0+}}{N_{0-}}+\frac{\Omega}{2}\ketbra{N_{+-}}{N_{0-}}+\text{H.c}\right]\notag\\-&(A_{xx}+\Delta)\ketbra{N_{0+}} -(A_{yy}+\Delta)\ketbra{N_{0-}} \notag\\
    -&A_{zz}\ketbra{N_{+-}},
\end{align}
where the first line contains the Zeeman and the MW terms, and the second and third lines contain the shifts. Note that a global $\Delta$ energy shift has been performed. An alternative way of understanding the bright and dark subspaces is by noticing that the bright subspace corresponds to the bosonic/ fully symmetric subspace, in which the states are symmetric under particle exchange. In turn, the dark subspace corresponds to the fermionic or antisymmetric subspace. Since the NV centers are considered to be parallel, and thus indistinguishable, they behave as bosons and we can only access the corresponding symmetric subspace if the system starts in the ground state.

Since we the two NV centers are usually initialized in state $\ket{00}$, we may restrict our analysis to the 6-level bright subspace. Moreover, by inspecting Fig.~\ref{fig:energy_diagram}(a) it becomes apparent that the states $\ket{P_{+-}}$ and $\ket{P_{0-}}$ can only be populated in the presence of a magnetic field. In the absence thereof, the relevant dimensionality of the system is further reduced to four levels. Among these four levels we identify two states of interest: $\ket{N}$ and $\ket{P}$ which are entangled states that belong to the double quantum transition. In the canonical 2 NV basis, these states read
\begin{align}
    \ket{N} = &\frac{1}{\sqrt{2}}(\ket{+1-1} + \ket{-1+1}),\\
    \ket{P} = &\frac{1}{\sqrt{2}}(\ket{+1+1} + \ket{-1-1}).
\end{align}
Note that $\ket{N}$ is insensitive to magnetic fields along the NV axes, since the first NV center accumulates a phase opposite to the second one. On the other hand, $\ket{P}$ features a fourfold increase in phase sensitivity compared to a single NV center in the single quantum transition, reaching the Heisenberg limit of measurement precision associated with the four electrons involved in the measurement. In turn, state $\ket{N}$ features a fourfold sensitivity to transverse electric fields and even though it cannot be used for magnetic field sensing it can be used to measure gradients of magnetic field. This makes them valuable states for sensing and we proceed to propose mechanisms for their direct production from the ground state. 

\section{\label{sec:results}Results and Discussion}
This section into three main parts, each focusing on the generation of a given entangled state belonging to the double quantum transition of the NV ground state. Subsections \ref{sec:N++} and \ref{sec:B++} make use of Raman transfer aided by $\ket{P_{0+}}$ to generate the entangled states. This is achieved by judicious combination of $\Omega$ and $\Delta$ values informed by an analysis based on adiabatic elimination of certain states from the diagram shown in Fig.~\ref{fig:energy_diagram}(a). Subsection \ref{sec:zero_mub} exploits the NV-ERC approach to generate an entangled state in zero-field conditions.

\subsection{\label{sec:N++}$\ket{N}$ state generation}
By inspecting Fig.~\ref{fig:energy_diagram}(a), Raman transfer  between $\ket{00}$ and $\ket{N}$ aided by the state $\ket{P_{0+}}$ may be achieved when the detuning of the $\ket{00}$ and the $\ket{N}$ states is the same, i.e $2\Delta = -\Azz$. This tentative result ignores the effect of the remaining levels and couplings. In a more precise analysis, adiabatic elimination of the states $\ket{P}, \ket{P_{0-}}$ and $\ket{P_{+-}}$ is used to find conditions on the values of the remaining parameters $\Omega, \mu B$ and $\Axx$ that enhance the effective shift of the $\ket{P_{0+}}$, thus improving the efficiency of the transfer. To implement the adiabatic elimination the Hamiltonian in Eq.~\eqref{eq:H_RWA_DARK} is divided in four 3x3 blocks 
\begin{equation}
\Ham = \mqty(\Ham_1 & \Ham_2\\ \Ham_2^\dagger & \Ham_3),
\end{equation}
where $\Ham_1$ contains the interactions between the three states of interest ($\ket{00}, \ket{N}$ and $\ket{P_{0+}}$), $\Ham_3$ contains the interactions between the eliminated states ($\ket{P}, \ket{P_{+-}}$ and $\ket{P_{0-}}$) and $\Ham_2$ contains the interactions between the kept and the eliminated states.
\begin{figure}
    \centering
    \includegraphics[width=\linewidth]{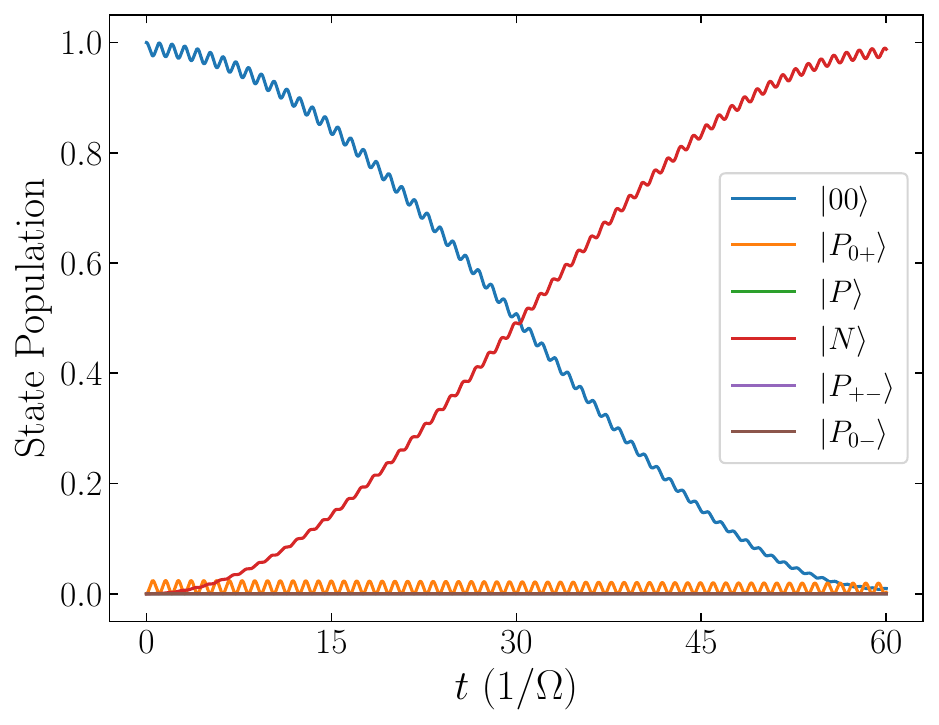}
    \caption{Raman transfer from state $\ket{00}$ to state $\ket{N}$ aided by the intermediate state $\ket{P_{0+}}$, with $\mu B = 0.05\Omega$, $\theta = 0.426\pi$, $(\mu_0\mu^2)/(4\pi r^3) = 10\Omega$ and $\Delta = -0.49875\Azz$.}
    \label{fig:Transfer_Dplus}
\end{figure}
These 3x3 matrices read
\begin{align}
    \Ham_1 =& \mqty(2\Delta& 0& \Omega/\sqrt{2}\\ 0 & -\Azz& \Omega/2\\ \Omega/\sqrt{2} & \Omega/2&\Axx+\Delta),\\
    \Ham_2 = & \mqty(0&0&0 \\ 0 &0 &0\\ \Omega/2&0&\mu B),\\
    \Ham_3 =& \mqty(\Azz&2\mu B&0\\ 2\mu B&\Azz&\Omega/2 \\ 0&\Omega/2&\Ayy+\Delta),
\end{align}

and the effective Hamiltonian $\Ham_{\text{eff}}$ resulting from adiabatic elimination is
\begin{equation}
    \Ham_{\text{eff}} = \Ham_1 - \Ham_2\Ham^{-1}_3\Ham^\dagger_2.
\end{equation}
Due to the particular shape of $\Ham_2$, the term $\Ham_2\Ham_3^{-1}\Ham_2^\dagger$ will only cause an additional shift $\delta$ in the state $\ket{P_{0+}}$. After performing adiabatic elimination the effective Hamiltonian reads
\begin{equation}\label{eq:Heff_N}
    \Ham_{\text{eff}} = \mqty(2\Delta& 0& \Omega/\sqrt{2}\\ 0 & -\Azz& \Omega/2 \\ \Omega/\sqrt{2} & \Omega/2 &\Axx+\Delta+\delta),
\end{equation}
\begin{figure}[h]
    \centering
    \includegraphics[width=\linewidth]{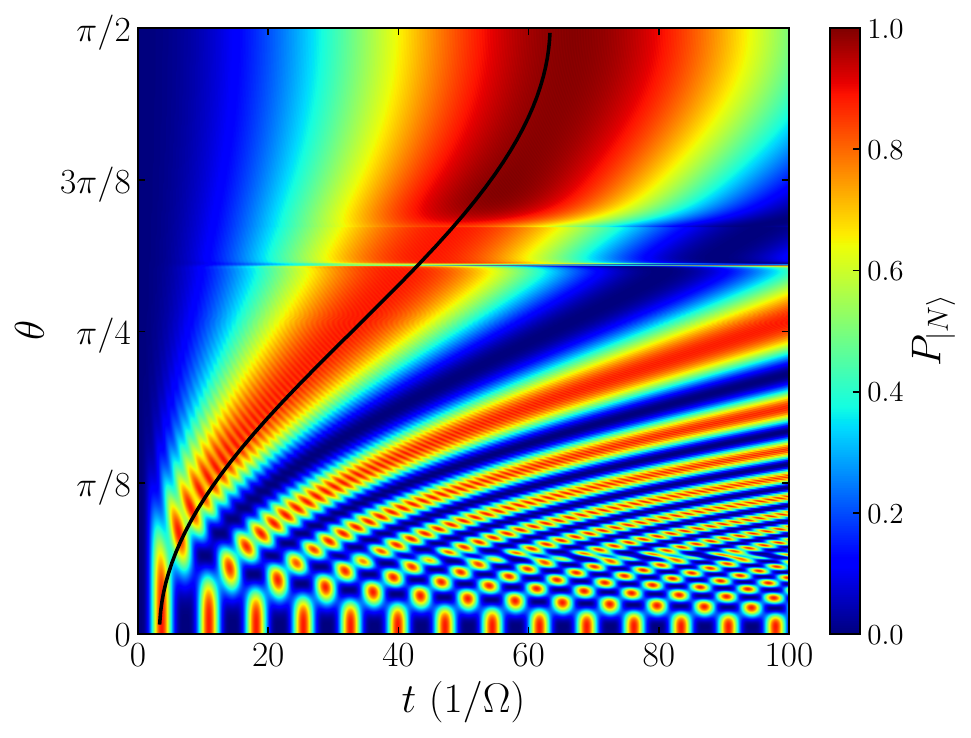}
    \caption{Time evolution of the population in the $\ket{N}$ state for $\theta\in[0,\pi/2]$ with $\mu B = 0.05\Omega$ and  $(\mu_0\mu^2)/(4\pi r^3) = 10\Omega$. The black line follows the maximum achieved population for each value of $\theta$.}
    \label{fig:Dplus_angle}
\end{figure}
\begin{figure}[H]
     \centering
     \includegraphics[width = \linewidth]{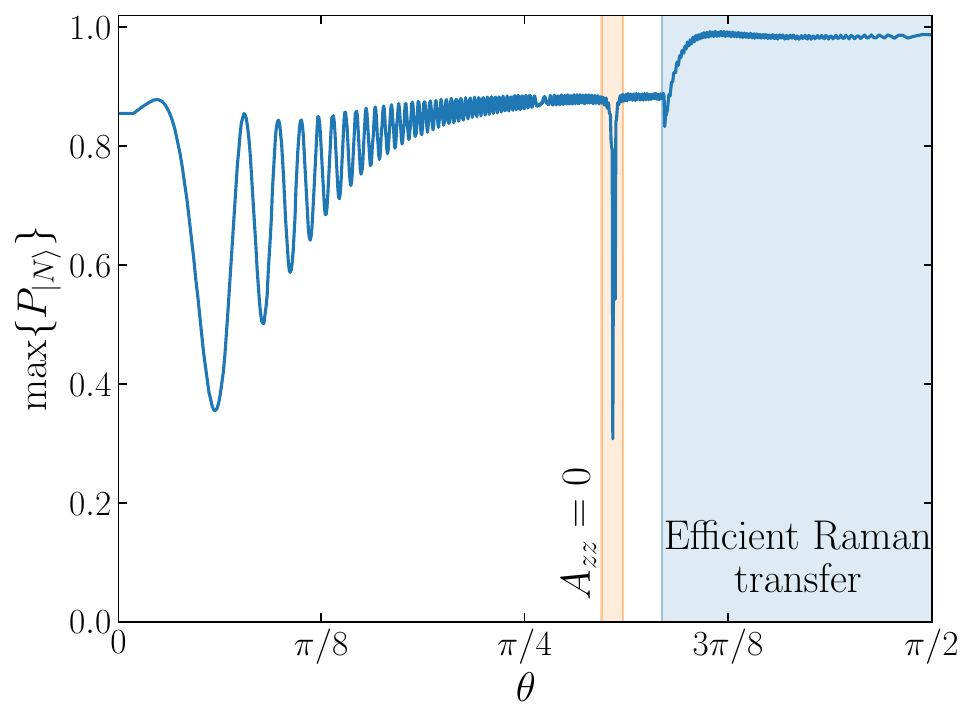}
     \caption{Maximum population in the $\ket{N}$ state depending on the angle $\theta$. The shaded region in orange corresponds to the region where \Azz vanishes and therefore the transfer is not achieved using $\mu B = 0.05\Omega$ and  $(\mu_0\mu^2)/(4\pi r^3) = 10\Omega$. The blue shaded region corresponds to the angle values for which an efficient Raman transfer is achieved.}
     \label{fig:population_Nplus}
 \end{figure}
where
\begin{widetext}
\begin{equation}\label{eq:delta}
    \delta = -\frac{(\mu B\Omega)^2\left[2(\Ayy+\Delta)-\Azz\right]}{4(\Ayy+\Delta)(\Azz-4(\mu B)^2)-\Azz\Omega^2}\left[\frac{2}{\Azz}+\frac{1}{\Ayy+\Delta}\right] - \frac{\Omega^2}{4\Azz}-\frac{(\mu B)^2}{\Ayy+\Delta}.
    \end{equation}
\end{widetext}
In the following, the conditions under which Raman transfer is most efficient are evaluated. In general, Raman transfer is achieved as long as $\Omega/2<\Axx+\Delta+\delta$. An increase in $\delta$ therefore allows for larger $\Omega$ and faster transfer. Although $\Delta\approx -\Azz/2$ is a good first guess, $\Delta$ must be finely tuned with the expression for $\delta$ in Eq.~\eqref{eq:delta}, as explained in Appendix \ref{sec:appendix_adiabatic}. An example of efficient $\ket{N}$ generation is depicted in Fig.~\ref{fig:Transfer_Dplus} which only negligibly populates the intermediate state.

To test the transfer in other parameter configurations, the angle $\theta$ that the unit vector connecting the two NV centers forms with the $z$ axis is considered. Since $\Axx\propto (1-3\sin^2\theta)$ and $\Azz\propto (1-3\cos^2\theta)$, it is only necessary to study the first quadrant, this is $\theta\in[0,\pi/2]$. The results are shown in Fig.~\ref{fig:Dplus_angle} and illustrate that the transfer is slower as $\theta$ grows.
\begin{figure}[H]
    \centering
    \includegraphics[width=\linewidth]{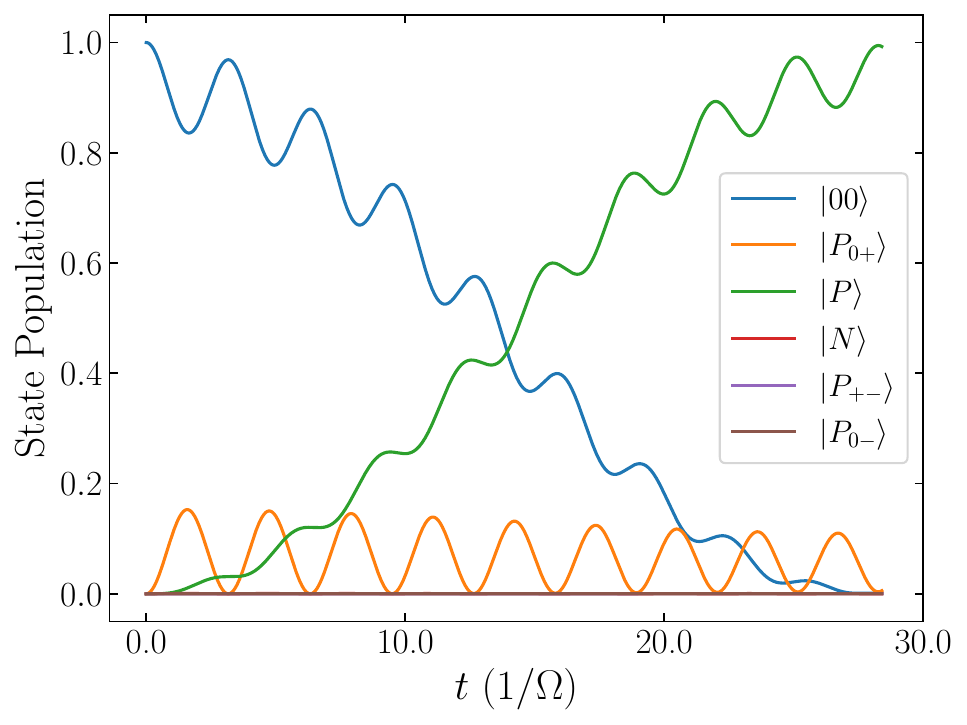}
    \caption{Raman transfer from state $\ket{00}$ to $\ket{P}$ aided by the intermediate state $\ket{P_{0+}}$ with $\mu B = 0.001\Omega$, $\theta = 0.292\pi$, $(\mu_0\mu^2)/(4\pi r^3) = 9.09\Omega$ and $\Delta = 0.504\Azz$.}
    \label{fig:Transfer_Bplus}
\end{figure}
When $\theta = 0$ the states $\ket{00}, \ket{P_{0+}}$ and $\ket{N}$ are resonant, allowing Raman transfer to happen faster. As $\theta$ grows, the detuning of the intermediate state increases, which decreases the transfer rate. An interesting feature takes place at $\theta = \arccos\left(1/\sqrt{3}\right)$, where the transfer is not achieved.

This is because, at that specific value, $\Azz = 0$ and not only $\ket{00}$ and $\ket{N}$, but also $\ket{P}$ becomes resonant, making it harder to obtain a high population on $\ket{N}$

due to population leakage into $\ket{P}$. The black line in Fig.~\ref{fig:Dplus_angle} indicates the time at which the maximum population in the state $\ket{N}$ is achieved for each angle. The population along this line is depicted in Fig. \ref{fig:population_Nplus}. It can be observed that for angles from $\theta = 0.334\pi$ to $\theta=\pi/2$ the transfer is much more effective. Moreover, when \Azz vanishes the population in state $\ket{N}$ drops significantly as a consequence of multiple resonances in the system that cause leakage to other states.
 
\subsection{\label{sec:B++}$\ket{P}$ state generation}
In analogy to the previous case, we consider the possibility of Raman transfer from $\ket{00}$ to $\ket{P}$ via $\ket{P_{0+}}$. A tentative choice to favor this mechanism is $2\Delta = \Azz$. However, this case is more subtle since the state of interest $\ket{P}$ is directly coupled via the external magnetic field to one of the states we aim to eliminate.

The relevant adiabatically elimination in this case is that of the states $\ket{P_{+-}}, \ket{P_{0-}}$ and $\ket{N}$ and the corresponding blocks $\Ham_1, \Ham_2$ and $\Ham_3$ read
\begin{equation}
        \Ham_1 =\mqty(2\Delta& 0& \Omega/\sqrt{2}\\ 0 & \Azz& \Omega/2\\ \Omega/\sqrt{2} & \Omega/2&\Axx+\Delta),
\end{equation}
\begin{equation}
        \Ham_2 = \mqty(0&0&0 \\ 2\mu B &0 &0\\ 0&\mu B&\Omega/2),
\end{equation}
\begin{figure}[H]
    \centering
    \includegraphics[width=\linewidth]{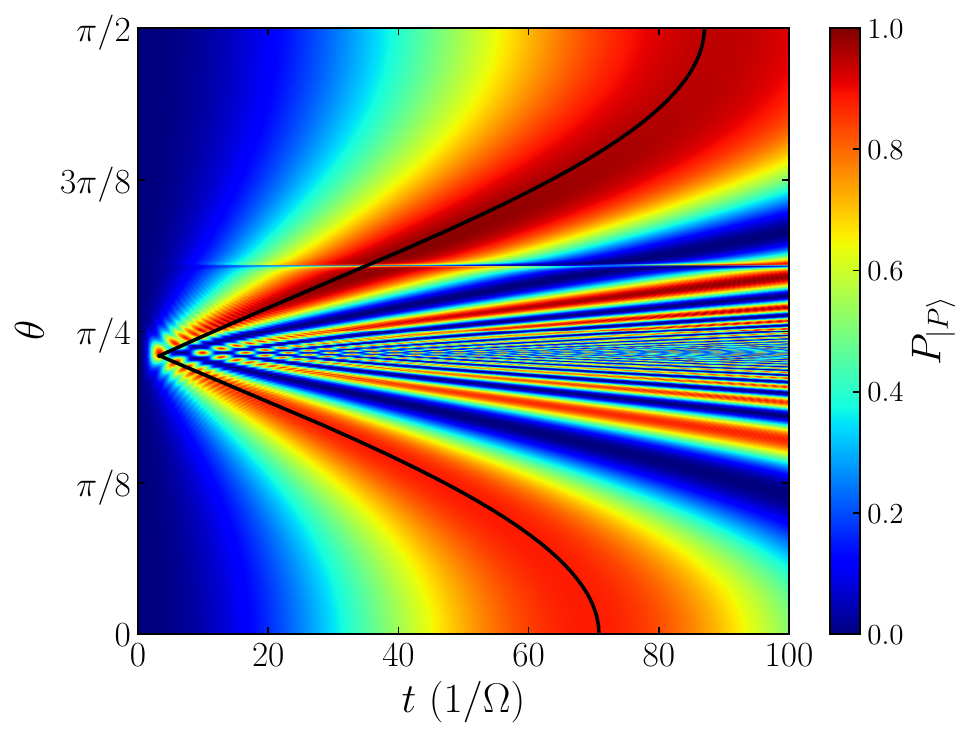}
    \caption{Time evolution of the population in the $\ket{P}$ state for $\theta\in[0,\pi/2]$ with $\mu B = 0.001\Omega$ and $(\mu_0\mu^2)/(4\pi r^3) = 9.091\Omega$. The black line follows the maximum achieved population for each value of $\theta$.}
    \label{fig:Bplus_angle}
\end{figure}
\begin{figure}[H]
    \centering
    \includegraphics[width=\linewidth]{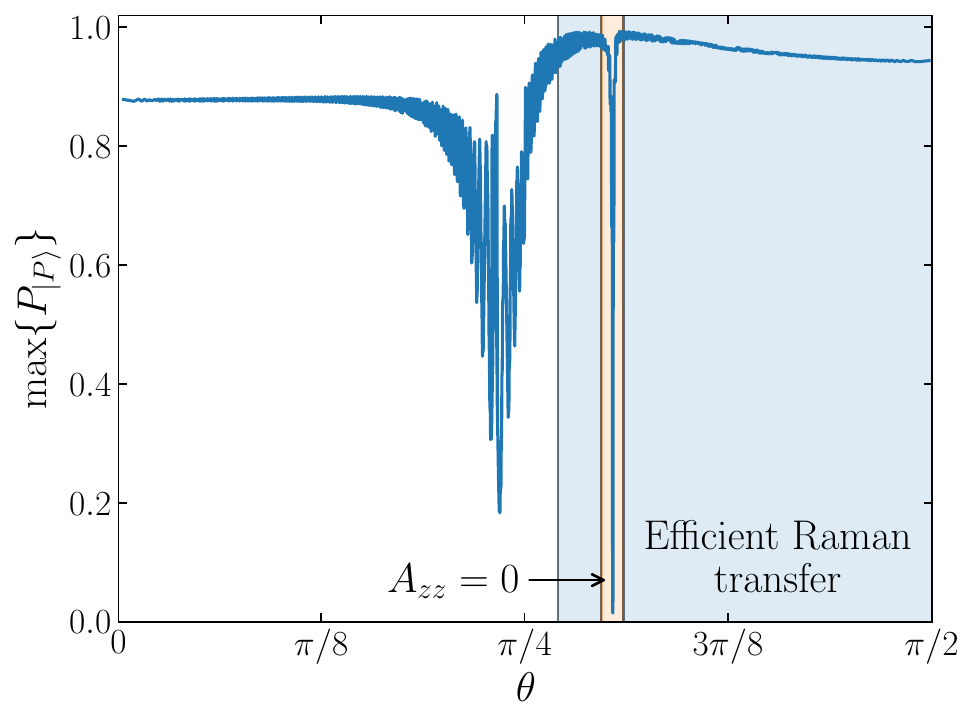}
    \caption{Maximum population in the $\ket{P}$ state depending on the angle $\theta$. The shaded region in blue corresponds to the region where an efficient Raman transfer is accomplished using $\mu B = 0.001\Omega$ and $(\mu_0\mu^2)/(4\pi r^3) = 9.091\Omega$. The shaded region in orange and signaled with an arrow corresponds to the region where \Azz vanishes and therefore the transfer is not achieved.}
    \label{fig:population_Pplus}
\end{figure}
\begin{equation}
        \Ham_3 =\mqty(\Azz&\Omega/2&0\\ \Omega/2&\Ayy+\Delta&0 \\ 0&0&-\Azz).
\end{equation}
In contrast with the previous case, there are more contributions in the effective Hamiltonian
\begin{equation}
    \Ham_{\text{eff}} = \mqty(2\Delta& 0& \Omega/\sqrt{2}\\ 0 & \Azz +\delta'& \Omega/2+\alpha\\ \Omega/\sqrt{2} & \Omega/2+\alpha&\Axx+\Delta+\delta''),
    \label{eq:Heff_P}
\end{equation}

where an additional coupling $\alpha$ between the states $\ket{P}$ and $\ket{P_{0+}}$, and additional shifts $\delta'$ and $\delta''$ appear after the adiabatic elimination
\begin{align}
    \delta' = & \frac{-4(\mu B)^2(\Ayy+\Delta)}{\Azz(\Ayy+\Delta) - \Omega^2/4},\\
    &\notag\\
    \alpha = & \frac{(\mu B)^2\Omega}{\Azz(\Ayy+\Delta)-\Omega^2/4},\\
    &\notag\\
    \delta'' = & \frac{-\Omega^2}{4\Azz} + \frac{(\mu B)^2}{\Ayy+\Delta-\Omega^2/4\Azz}.
\end{align}

To achieve an efficient transfer several conditions need to be satisfied. First of all, the value of $\alpha$ has to be very small. Moreover, the effective detuning $\Axx+\Delta+\delta'$ needs to be large. Lastly, the detunings from states $\ket{00}$ and $\ket{P}$ need to be very close, if not equal. In order to fulfill all these three conditions, low magnetic field $B$ and pulse strength $\Omega$ are required. Low magnetic field is required since $\ket{P}$ and $\ket{P_{+-}}$ are coupled with $2\mu B$ and as the magnetic field grows, there is a population leakage to $\ket{P_{+-}}$ that prevents the generation of $\ket{P}$. Low pulse strength is required as per the Raman transfer condition. Therefore, the hierarchy in the parameters in this regime is $\mu B\ll\Omega<\Azz, \Axx$.

Having set this hierarchy, note that both $\alpha$ and $\delta'$ tend to 0. Therefore, $\Delta\approx \Azz/2$ a good condition for transfer. An instance of Raman transfer to $\ket{P}$ is shown in Fig.~\ref{fig:Transfer_Bplus}. Although the population is completely transferred from the ground state to the $\ket{P}$ state, the process is not as smooth as in the example in Fig.~\ref{fig:Transfer_Dplus}. In fact, in this case, the population in the intermediate state oscillates with a significant amplitude. This shows the importance of finely tuning the parameters. However, it is also worth noticing that in this case the entangled state $\ket{P}$ is generated much faster than $\ket{N}$, which establishes a trade-off between the smoothness of the Raman transfer and its rate.

The transfer to the $\ket{P}$ state as a function of $\theta$ is shown in Fig.~\ref{fig:Bplus_angle}. As in Fig.~\ref{fig:Dplus_angle} the same effect at $\theta = \arccos(1/\sqrt{3})$ appears, where $\Azz$ vanishes, preventing the transfer of all the population to $\ket{P}$. For $\theta$ values between $\theta = \arcsin(1/\sqrt{3})$ and $\theta = \arccos(1/\sqrt{3})$, (the points where $\Axx$ and $\Azz$ vanish, respectively) we find higher frequency rates for the Raman transfer. As in the previous case, it can be observed that there is a region where the transfer achieved is much more efficient, which corresponds to angles above $\pi/4$. Fig.~\ref{fig:population_Pplus} illustrates the maximum population achieved in $\ket{P}$ as a function of $\theta$. The transfer is very efficient for $\theta >\pi/4$, except when \Azz vanishes. For small values of $\theta$ the transfer is not very efficient in terms of population. However, as a consequence of being far from resonance, the maximum population reached is almost constant.

Additionally, another advantage of the direct generation of the states $\ket{P}$ and $\ket{N}$, without the participation of intermediate states, is that the generation time can be effectively counted as sensing time, unlike conventional approaches, where the time spent generating the states does not contribute to the sensing process. Therefore, bypassing intermediate states enhances temporal efficiency in quantum sensing applications.

\subsection{Comparison with other entanglement generation proposals}
\begin{table*}[htbp]
\centering
\caption{\label{tab:entangled}Comparison between different NV-based entanglement methods. The table contains the method used to generate entangled states between two or more NV centers, the fidelity achieved, whether the Heisenberg limit (HL) is fulfilled and whether the NV centers are parallel or not. The two highlighted rows in blue correspond to the proposal in this manuscript and the rows highlighted in orange correspond to experimental demonstrations of entanglement between NV centers. Remaining rows are theoretical proposals.}
\begin{tabular}{@{}l@{}c@{}c@{}c@{}c@{}}\hline\hline
& Method used & \hspace{.6cm}Fidelity\hspace{.6cm} & \hspace{.6cm}HL\hspace{.6cm} & \hspace{.6cm}Parallel NV centers\\ \hline
\rowcolor{light_blue}State $\ket{P}$ & Raman Transfer & 0.99735 & \checkmark & \checkmark\\
\rowcolor{light_blue}State $\ket{N}$ & Raman Transfer & 0.99972 & \checkmark & \checkmark\\
Phys. Rev. Lett. \textbf{110}, 156402 \cite{hybrid_1}\hspace{0.6cm} & Spin Squeezing & - & \xmark & \checkmark\\
Phys. Rev. X \textbf{3}, 041023 \cite{hybrid_2} & Ferromagnet & 0.99976 & \xmark & \checkmark\\
Scientific Reports \textbf{12}, 14964 \cite{Annealing_Matsuzaki} & Quantum Annealing &0.979 & \xmark & \checkmark\\
\rowcolor{light_orange}Phys. Rev. X \textbf{15}, 021069 \cite{Fedor_Entanglement} & Pulse Sequence & 0.96 & \xmark & \xmark\\
\rowcolor{light_orange}Nature Physics \textbf{9}, 139-143 \cite{Dolde2013} & Pulse Sequence & 0.67 & \xmark & \xmark\\
\hline\hline
\end{tabular}
\end{table*}
In the following we compare our proposal with already existing methods to generate entangled states between NV center pairs (see Table.~\ref{tab:entangled}).  It is important to emphasize that we propose the generation of entangled states from the double quantum transition directly from the ground state, without populating intermediate states. These states achieve the Heisenberg limit, making them suitable for entanglement-enhanced sensing purposes, specifically longitudinal magnetic fields and transverse electric fields. We report maximum fidelities of 0.99735 for state $\ket{P}$ and 0.99972 for state $\ket{N}$.

The works \cite{hybrid_1, hybrid_2} make use of hybrid systems to generate entangled states in parallel NV center setups. The results in \cite{hybrid_1} concern entanglement generation based on spin squeezing of an ensemble of parallel NV centers. While the squeezing parameter scaling shows the potential for entanglement-enhanced magnetometry, the Heisenberg limit is not reached. In \cite{hybrid_2}, only the single quantum transition is considered, so the Heisenberg limit cannot be reached in this context. The two spins are coupled using a ferromagnetic coupler, enabling long-distance entanglement. Nevertheless, the fidelity obtained of the entangling gate reaches 0.99976, slightly higher than the fidelity obtained for state $\ket{N}$ in this manuscript.

The proposal displayed in \cite{Annealing_Matsuzaki} relies on a symmetry that is only preserved under exactly zero longitudinal magnetic field. The protocol abruptly breaks down in the presence of even the slightest magnetic field, rendering it impractical for sensing purposes. Assuming the possibility to operate under these stringent conditions, they report generation of state $\ket{P}$ with a fidelity of 0.979, performing well below our protocol. Furthermore, quantum annealing necessarily populates intermediate states, which can result in the collection of undesired perturbations.

\subsection{Zero-field entanglement generation}\label{sec:zero_mub}

\begin{figure}[H]
    \centering
    \includegraphics[trim = 2.8cm 11cm 36.8cm 3.2cm, clip=True, width=0.7\linewidth]{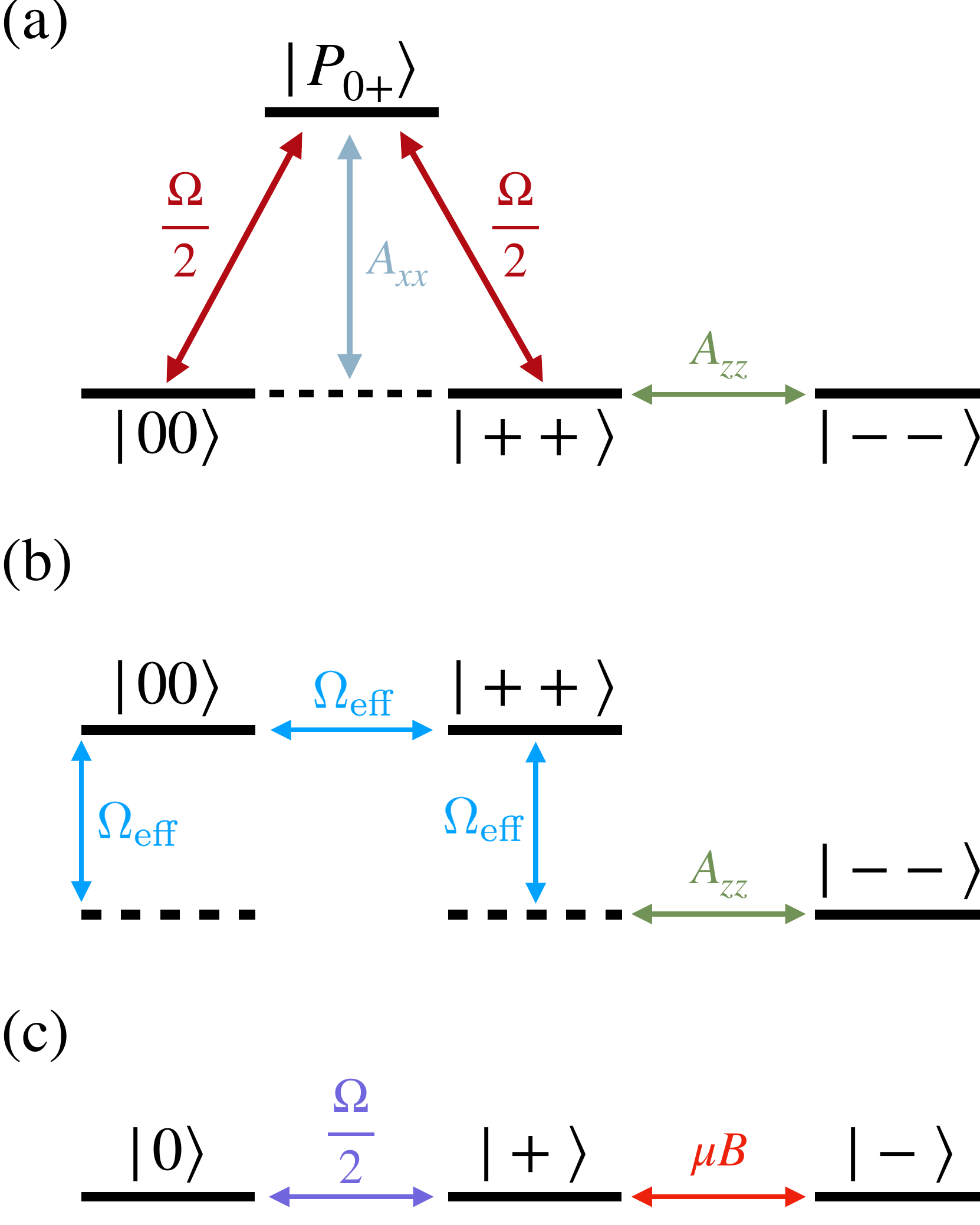}
    \caption{(a) States of the system in zero field conditions after the change of basis. In wine red the couplings coming the MW field. In olive green the coupling coming from the $\Sz\otimes\Sz$ interactions and in gray the shift coming from the $\Sx\otimes\Sx$ interaction. (b) States after the adiabatic elimination is performed. In blue, the effective coupling. (c) NV-ERC scheme with states $\ket{0}$ and $\ket{+}$ shifted by $\Omega /2$ relative to state $\ket{-}$.}
    \label{fig:diagrams_noB}
\end{figure}

As it has been discussed in Sec.~\ref{sec:preliminaries}, in the absence of a bias magnetic field, the dynamics of the system are reduced to a four-level system [see Fig.~\ref{fig:energy_diagram}(a)], allowing to apply almost directly the NV-ERC technique \cite{Paper_Alberto}. By performing a change of basis to $\lbrace\ket{00}, \ket{P_{0+}}, \ket{++},\ket{--}\rbrace$, the level scheme shown in Fig.~\ref{fig:diagrams_noB}(a) is obtained. In this setting, a maximally entangled state in the double quantum transition corresponds to an equal superposition of states $\ket{++}$ and $\ket{--}$, up to a relative phase. 
By adiabatically eliminating the intermediate state $\ket{P_{0+}}$, the insight gained with NV-ERC may assist in identifying parameters that achieve this goal.

The three blocks from the adiabatic elimination $\Ham_1, \Ham_2$ $\Ham_3$ and the corresponding effective Hamiltonian read 

\begin{align}
    \Ham_1 =& \mqty(0 & 0 & 0\\ 0 & 0 & A_{zz}\\ 0 & A_{zz} & 0), \\
    \Ham_2 = & \mqty(\Omega /2 & \Omega /2 & 0)^T,\\
    \Ham_3 =& \Axx,\\
    \Ham_{\text{eff}} =& \mqty(\Omega_{\text{eff}} & \Omega_{\text{eff}} & 0\\ \Omega_{\text{eff}} & \Omega_{\text{eff}} &\Azz\\ 0 & \Azz & 0),\label{eq:H3}
\end{align}
revealing an effective coupling $\Omega_{\text{eff}} = -\Omega ^{2}/(4A_{xx})$ between states $\ket{00}$\ and $\ket{++}$ [see Fig.~\ref{fig:diagrams_noB}(b)].

An analogy between this system and the original NV-ERC scheme can be established by mapping the NV-ERC parameters $\mu B \to \Azz$ and $\Omega/2 \to \Omega_{\text{eff}}$ [see Fig.~\ref{fig:diagrams_noB}(c)]. In contrast to NV-ERC, states $\ket{00}$ and $\ket{++}$ are shifted $\Omega_{\text{eff}}$ with respect to $\ket{--}$. However, this conserves the crucial feature of NV-ERC, which consists in the ability to fully depopulate the state $\ket{00}$ and achieve a superposition state in the equator of the double quantum transition. We proceed to exploit this feature for the generation of an equal superposition of $\ket{++}$ and $\ket{--}$, i.e. an entangled state.

Due to the fact that NV-ERC addresses the zero-field line (i.e the MW driving frequency $\omega =D$) the detuning $\Delta = \omega-D$ vanishes. Additionally, the set of parameters that allow operation with this approach must satisfy the two regimes imposed by NV-ERC and Raman transfer, $\Azz < \Omega^2/4\Axx$ and $\Omega/2 < \Axx$, respectively. We find that this regime is satisfied for $\theta \simeq \arccos(1/\sqrt{3})$. Additionally, with insight from NV-ERC, we also find that $\Omega_{\text{eff}} = 1.293 \Azz$ in the 3-level model Eq.~\eqref{eq:H3} produces an equal superposition of states $\ket{++}$ and $\ket{--}$ while simultaneously depopulating state $\ket{00}$. 

\begin{figure}
    \centering
    \includegraphics[width=\linewidth]{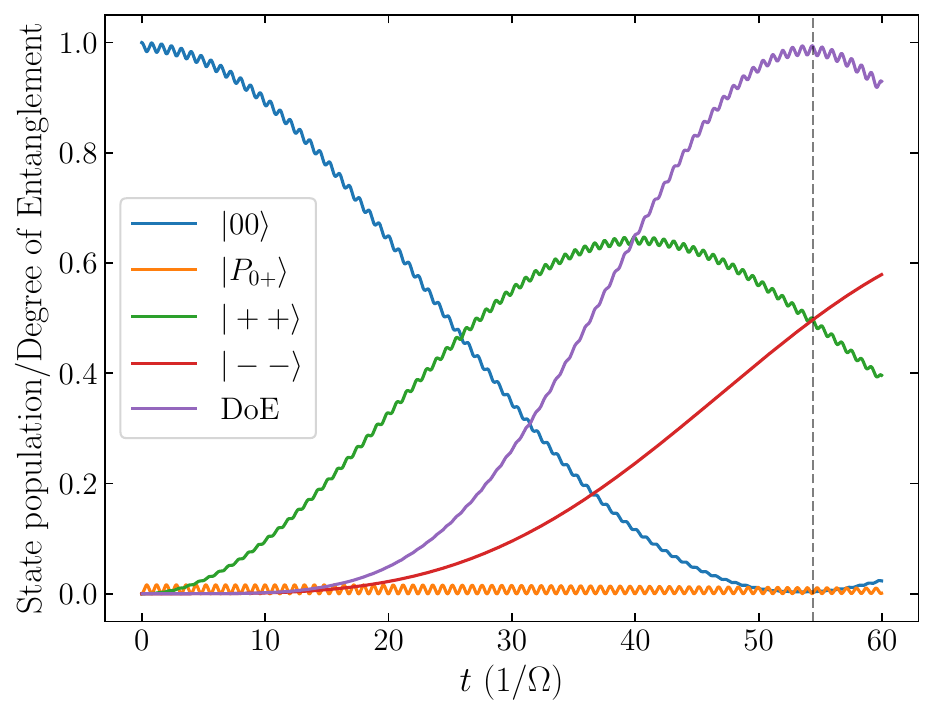}
    \caption{Evolution of the four states in the absence of an external magnetic field. The vertical dashed line signals the time where the system reaches the maximally entangled state $\ket{\Psi}$ as the ground state becomes unpopulated. The $y$ axis represents the population of each state and the degree of entanglement (DoE) of the evolved state which ranges from 0 for a separable state to 1 for a maximally entangled state. $\theta = 0.303\pi, \Omega = 40\Azz, \Axx = \Omega^2/(4\times 0.1293\Azz) $ have been used for this setup.}
    \label{fig:entanglement_mub0}
\end{figure}

This prediction transfers smoothly to the full four-level Hamiltonian as illustrated in Fig.~\ref{fig:entanglement_mub0}, where we show how, starting from the ground state $\ket{00}$, a MW pulse at frequency $D$ produces the maximally entangled state $\ket{\Psi} \simeq \frac{1}{\sqrt{2}}(\ket{++} + e^{-i\pi/4}\ket{--})$. This result provides evidence of the existence of an avenue of exploration for the design of alternative and increasingly optimal strategies for the generation of Heisenberg-limited sensing states for parallel NV center pairs.

\subsection{Implementation challenges}
The experimental implementation of the proposed approach faces two main challenges. Firstly, hyperfine splitting due to interactions with the nitrogen nuclear spin introduces instability in the bias magnetic field. In each experimental realization, it will take one of two or three possible values depending on the state and isotopic value of the nucleus. Polarization of the nucleus or a postselection scheme depending on its state can mitigate this effect.

Secondly, the generation of entangled states is inherently slow due to weak dipole-dipole interactions. This challenge is not unique to our approach but affects all NV center-based entanglement-generation proposals. While stronger dipole-dipole interactions necessitate closer NV centers, distances below 3 nm result in exchange interactions surpassing dipole-dipole interactions \cite{PhysRevB.93.220402}, causing the NV centers to lose their distinct properties. Beyond this distance, the dipole interaction rapidly decays to the order of kHz, leading to entanglement generation rates in the range of milliseconds. Consequently, it is difficult to produce states within the $T_2$ decoherence time of the system. A recent attempt at enhancing the dipole-dipole interaction strength by means of ferromagnetic couplers may constitute a solution to this challenge \cite{hybrid_2}.

Moreover, fabricating samples with two parallel, dipole-coupled NV centers is more complex than manufacturing samples with single NV centers. Nonetheless, recent advancements in NV center manufacturing have shown promising results in increasing the availability of coupled NV center samples \cite{BALASUBRAMANIAN2022282,Liang2025}.\\

Additionally, in laboratory conditions, the system at study will be under the influence of external noise sources, such as thermal fluctuations, laser instability or MW field imperfections, among others. Laser instability will affect the protocol similarly to any NV center-based setup, since it is only used for initialization in the $\ket{00}$ state and for readout.

Thermal fluctuations induce a shift of the zero-field splitting, which can result in a slight instability when tuning the driving frequency to match the optimal detuning for an efficient Raman transfer. Furthermore, phonon-induced decoherence limits the operation time. However, this noise source can be mitigated to some extent by employing cryostats that maintain a stable temperature.

MW field imperfections can give rise to errors in the performance of quantum gates, which can result in a reduction on the fidelities of target states and can lead to decoherence. This challenge is mitigated by the fact that our proposal does not require the use of several MW pulses or changes in the frequency of the driving field during the protocol. Moreover, the magnitude of the error will depend on the angle the NV centers subtend. As it is shown in Fig.~\ref{fig:Dplus_angle} and \ref{fig:Bplus_angle}, the protocol is robust to timing errors when the Raman transfer is slow. In this regime, phase or timing errors in the MW field will not cause significant imperfections in the protocol. However, for $\theta$ values corresponding to fast transfer regimes, such errors may prevent the protocol from working as intended. This source of noise can be mitigated by using calibrated high-precision MW sources.

\section{\label{sec:conclusions}Conclusions}
We propose a novel mechanism based on the NV-ERC approach that uses globally addressing MW pulses on two dipole-coupled parallel NV centers to prepare maximally entangled states of the double quantum transition of the NV ground state.

After simplifying the system by performing an appropriate change of basis and moving to a rotating frame, we have found the conditions that allow to generate the desired states via Raman transfer, assessed by an analysis based on adiabatic elimination. Afterwards, we have studied the behaviour of the Raman transfer with the angle $\theta$ that the two NV centers subtend, concluding that for $\theta$ values close to the resonance between the states in which we are interested, the transfer becomes faster. We have proposed an additional scheme in the case of vanishing bias magnetic field.

The generated states can serve multiple purposes, becoming sensors of transverse electric fields or longitudinal magnetic fields. Because they belong to the double quantum transition, they feature a fourfold improved sensitivity when compared to a single NV center implementation, reaching the Heisenberg limit.

\begin{acknowledgments}
J.C. acknowledges support from grant PID2021-124965NB-C22 funded by MICIU/AEI/10.13039/501100011033 and by ``ERDF/EU''.
M.M.R, A.L.G. and J.C. acknowledge support from European Union project C-QuENS (Grant No. 101135359). J.C. additionally acknowledges support from grant CNS2023-144994 funded by MICIU/AEI/10.13039/501100011033 and by ``ERDF/EU''.
\end{acknowledgments}

\bibliographystyle{unsrtnat}
\bibliography{references}

\newpage
\appendix
\section{Dipole-dipole interaction Hamiltonian}\label{sec:appendix_dipole}
In this section we derive step by step the dipole-dipole interaction Hamiltonian in Eq.~\eqref{eq:H_dip} after transforming into the rotating frame with respect to $\Ham_0 = \omega\left(\hat{S}^2_{1,z}+\hat{S}^2_{2,z}\right)$ and applying the RWA. We consider two NV centers in the $x-z$ plane, with their axes parallel, thus being spectrally indistinguishable. The two NV centers are separated a distance $r$ and the unit vector connecting them is given by $\hat{\mathbf{r}} = (\sin\theta, 0, \cos\theta)^T$ in spherical coordinates. When expanding the products in the Hamiltonian in Eq.~\eqref{eq:H_dip}, we find that it can be expressed in the form
\begin{equation}
    \Ham_{\text{dip}} = \sum_{jk}A_{jk}\hat{S}_j\otimes \hat{S}_k,
\end{equation}
with $j,k = x,y,z$ and
\begin{equation}
    A_{jk} = \frac{\mu_0\mu^2}{4\pi r^3}\left[\delta_{jk} - 3\hat{\mathbf{r}}_j\hat{\mathbf{r}}_k\right],
\end{equation}
where $\delta_{jk}$ is Kronecker's delta. We analyze how all the different terms in the Hamiltonian read in the interaction picture separately and we apply the RWA. Note that in order to simplify the notation this derivation is performed in the $\lbrace\ket{+},\ket{0},\ket{-}\rbrace^{\otimes 2}$ basis.
\begin{itemize}
    \item $\Azz\,\Sz\otimes\Sz$ term: Due to the fact that $\commutator{\Ham_0}{\Sz\otimes\Sz} = 0$, after going to the interaction picture, no phase is collected by this term, leaving it unchanged.
    \item $A_{yk}\,\Sy\otimes\hat{S}_k$ and $A_{ky}\,\hat{S}_k\otimes\Sy$ terms with $y\neq k$: Even before going into the interaction picture these terms are 0 since $y\neq k$ and $\hat{\mathbf{r}}_y = 0$.
    \item $A_{xz}\Sx\otimes\Sz$ and $A_{zx}\Sz\otimes\Sx$ terms: \Sx can be rewritten as
    \begin{equation}\label{eq:Sx}
        \Sx = \ketbra{0}{+} + \text{H.c},
    \end{equation}
    where $\ket{\pm} = \frac{1}{\sqrt{2}}(\ket{1}\pm\ket{-1})$. In order to go to the interaction picture, each NV center is transformed with the unitary $\hat{U} = e^{i\omega t\Sz^2}$ as
    \begin{equation}
    \hat{U}\Sx\hat{U}^\dagger \otimes \hat{U}\Sz\hat{U}^\dagger = (e^{-i\omega t}\ketbra{0}{+} + \text{H.c})\otimes\Sz,
    \end{equation}
    which results in \Sx collecting a phase. If this tensor product is expanded, all the terms contain temporal dependences. When RWA is applied all these terms vanish, yielding no contributions from these terms after applying the RWA.
    \item $\Axx\Sx\otimes\Sx$ term: \Sx is expressed as in Eq.~\eqref{eq:Sx}. Upon moving to the interaction picture, the $\Sx\otimes\Sx$ term reads
    \begin{align}
    \hat{U}\Sx\hat{U}^\dagger \otimes \hat{U}\Sx\hat{U}^\dagger &= (e^{-i\omega t}\ketbra{0}{+} + \text{H.c})^{\otimes 2} \notag\\
    & = e^{-2i\omega t}\ketbra{00}{++} + \ketbra{0+}{+0} + \text{H.c}.
    \end{align}

    Applying the RWA to this expression, the terms with $e^{\pm 2i\omega t}$ vanish and the remaining terms read
    \begin{equation}
    \Axx\left[\hat{U}(\Sx\otimes\Sx)\hat{U}^\dagger\right] \overset{\text{RWA}} {\longrightarrow}\Axx\left(\ketbra{0+}{+0} + \text{H.c}\right).
\end{equation}
    \item $\Ayy\Sy\otimes\Sy$ term: \Sy is expressed as
    \begin{equation}
        \Sy = i\ketbra{0}{-} + \text{H.c}.
    \end{equation}

    In a similar way as with the previous term, when moving to the interaction picture, the $\Sy\otimes\Sy$ term reads
    \begin{align}
\hat{U}\Sy\hat{U}^\dagger \otimes \hat{U}\Sy\hat{U}^\dagger &= (ie^{-i\omega t}\ketbra{0}{-} + \text{H.c})^{\otimes 2} \notag\\
    & = -e^{-2i\omega t}\ketbra{00}{--} + \ketbra{0-}{-0} + \text{H.c}.
    \end{align}
\end{itemize}

After performing the RWA, the terms containing $e^{\pm2i\omega t}$ vanish and the remaining terms read
\begin{equation}
    \Ayy\left[\hat{U}(\Sy\otimes\Sy)\hat{U}^\dagger\right] \overset{\text{RWA}} {\longrightarrow}\Ayy\left(\ketbra{0-}{-0} + \text{H.c}\right).
\end{equation}

There are only contributions from the $\Sx\otimes\Sx, \Sy\otimes\Sy$ and $\Sz\otimes\Sz$ terms. If all the different contributions are added up, the dipole-dipole interaction Hamiltonian in the rotating frame after the RWA reads
\begin{align}
    \Ham_{\text{dip, RWA}} & = \Axx\ketbra{0+}{+0} + \Ayy\ketbra{0-}{-0}\notag\\
    &+\Azz\ketbra{++}{--} + \text{H.c}.
\end{align}

\section{Fine tuning of $\Delta$}\label{sec:appendix_adiabatic}
To generate the $\ket{N}$ state, starting from Eq.~\eqref{eq:Heff_N}, if an additional adiabatic elimination of state $\ket{P_{0+}}$ is performed, the condition to have a resonant Raman transfer reads
\begin{equation}\label{eq:poly_N}
    2\Delta = -\Azz + \frac{\Omega^2}{4(\Axx + \Delta+\delta)}.
\end{equation}
When the expression for $\delta$ is substituted here, it becomes a third-degree polynomial in $\Delta$ which must be solved numerically.\\

The third-degree polynomial in $\Delta$ can be expressed as $a\Delta^3 + b\Delta^2+c\Delta+d=0$, where $a,b,c,d$ depend on $\Omega, \Axx, \Ayy, \Azz$ and $\mu B$. The change of the coefficients with slight variations of the parameters is discussed below. When slightly varying $\Omega$, i.e $\Omega\to\Omega+\varepsilon$, all the coefficients of the polynomial resulting from Eq.~\eqref{eq:poly_N} change except the coefficient of $\Delta^3$. Up to first order in $\varepsilon$, the change in coefficients is

\begin{align}
        b\to& b-\Omega\varepsilon,\\
        c\to& c-\left(\Axx+\frac{5}{2}\Azz\right)\Omega\varepsilon,\\
        d\to& d+\left(\frac{3}{2}\Axx\Azz+\frac{\Azz^2}{2}\right)\Omega\varepsilon.
    \end{align}
When slightly varying $\mu B$ $(\mu B\to \mu B+\varepsilon)$, a similar behavior is observed, with all the coefficients varying linearly with $\mu B\varepsilon$ except the coefficient with $\Delta^3$, which remains unchanged, up to first order in $\varepsilon$.

In the case of the generation of the $\ket{P}$ state, performing the additional adiabatic elimination of $\ket{P_{0+}}$ in Eq.(\ref{eq:Heff_P}) yields the following resonant Raman transfer condition
\begin{equation}\label{eq:poly_P}
    2\Delta - \frac{\Omega^2}{2(\Axx + \Delta+\delta'')} = \Azz+\delta'-\frac{\left(\frac{\Omega}{2} + \alpha\right)^2}{\Axx+\Delta+\delta''},
\end{equation}
which again yields a third degree polynomial in $\Delta$ that has to be numerically solved after the expressions for $\delta'$, $\delta''$ and $\alpha$ are substituted. 

When varying $\mu B$ in the polynomial resulting from Eq.~\eqref{eq:poly_P}, the same behavior is observed. However, when slight variations of $\Omega$ are considered the change of the coefficients is slightly different. The cubic coefficient remains unchanged, the quadratic and the linear term change proportional to $\Omega\varepsilon$, but the independent term changes as
\begin{equation}
        d\to d+[12\Azz^2(\Axx+\Azz) + 4\Azz^2\Axx]\Omega\varepsilon + 2\Azz\Omega^3\varepsilon,
    \end{equation}
where there is an additional term proportional to a higher power of $\Omega$. Numerically, the influence of variations in $\mu B$ limited to $\ket{P}$ generation. However, since $\mu B \ll \Omega$ in the low-field regime considered, such variations in $\mu B$ do not significantly affect the solutions of the polynomial.

The parameters \Axx and \Azz, which depend on the angle $\theta$, have been varied continuously from 0 to $\pi/2$. The numerical solutions for $\Delta$ vary smoothly with the parameters and always remain very close to $\pm\Azz/2$, for $\ket{P}$ and $\ket{N}$ respectively. As for \Ayy, its impact is not treated separately since it can be expressed in terms of \Axx and \Azz.

\begin{figure}[H]
        \centering
        \includegraphics[width=\linewidth]{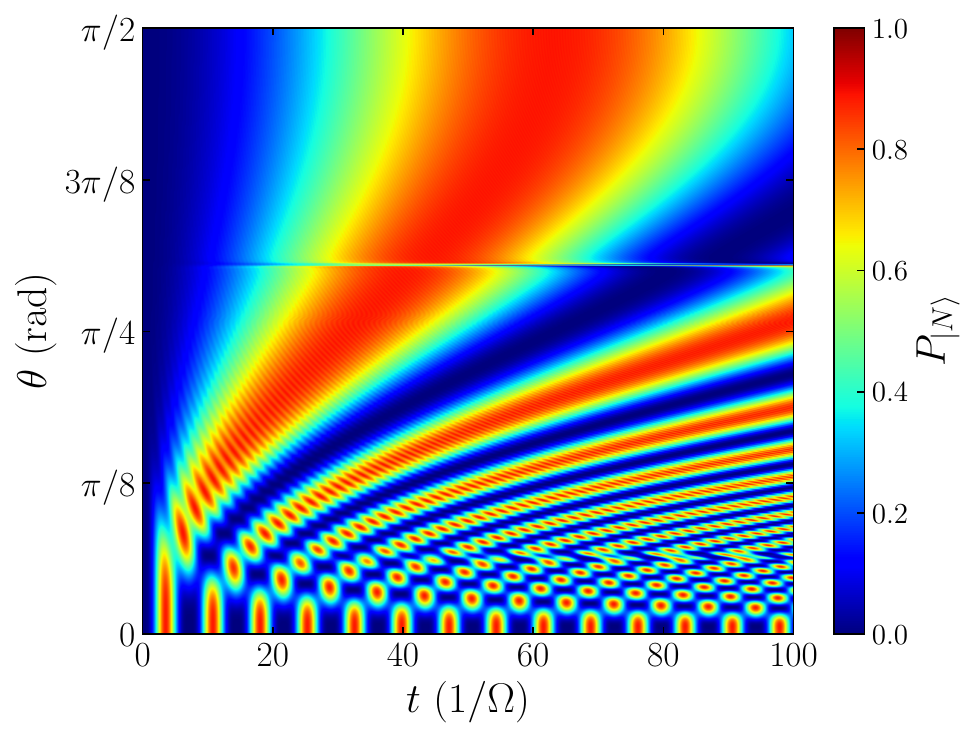}
        \caption{Time evolution of the population in the $\ket{N}$ state for $\theta\in[0,\pi/2]$ with $\mu B = 0.05\Omega$,  $(\mu_0\mu^2)/(4\pi r^3) = 10\Omega$ and $\Delta=-\Azz/2$. The black line follows the maximum achieved population for each value of $\theta$.}
        \label{fig:transfer_detuning_mal_N}
    \end{figure}

As long as the detuning of the MW drive is close to $\pm \Azz/2$ for the generation of the states $\ket{P}$ and $\ket{N}$ respectively, a good transfer will be obtained. However, it will not be optimal since the effects introduced when performing the adiabatic elimination will not be taken into account. In Fig.~\ref{fig:transfer_detuning_mal_N} the Raman transfer when choosing $\Delta = -\Azz/2$ is shown. This situation can be interpreted as obtaining a $\Delta$ value for a given set of parameters, different from those that optimize the protocol. With this change of parameters there is not a significant change on the behavior of the system, except for the maximum populations achieved, which are not as good as when the $\Delta$ value corresponds to the solution of the third degree polynomial.

\begin{figure}[H]
        \centering
        \includegraphics[width=\linewidth]{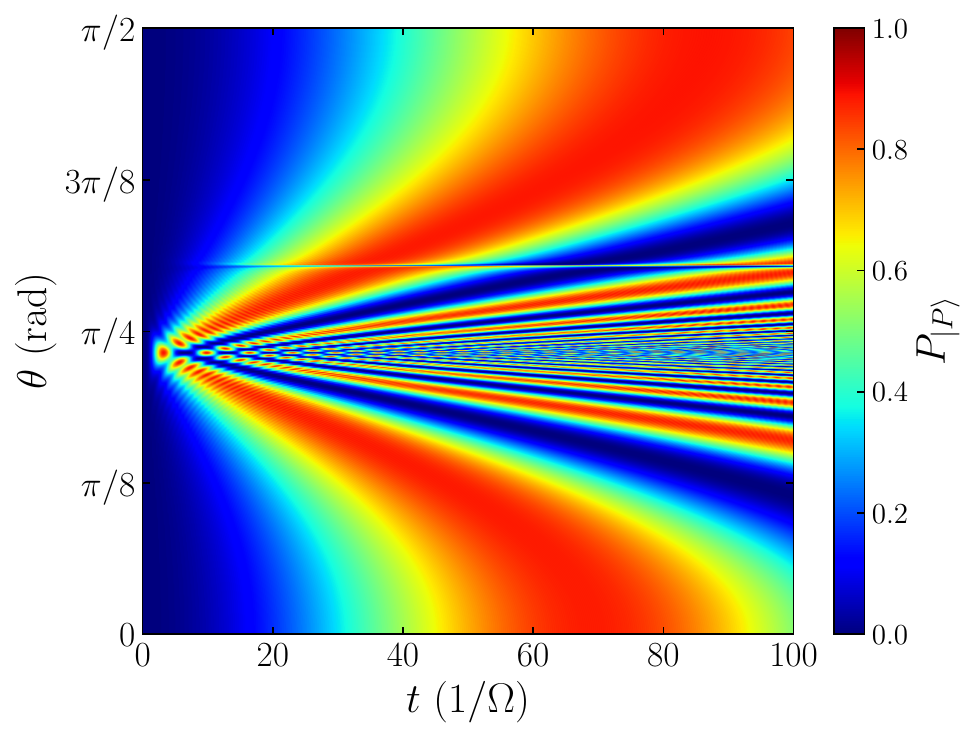}
        \caption{Time evolution of the population in the $\ket{P}$ state for $\theta\in[0,\pi/2]$ with $\mu B = 0.001\Omega$, $(\mu_0\mu^2)/(4\pi r^3) = 9.091\Omega$ and $\Delta=\Azz/2$. The black line follows the maximum achieved population for each value of $\theta$.}
        \label{fig:transfer_detuning_mal_P}
    \end{figure}

The same happens in Fig.~\ref{fig:transfer_detuning_mal_P}, where $\Delta = \Azz/2$ has been chosen to generate $\ket{P}$. From these two figures it can be interpreted that as long as $\Delta$ is close to $\pm\Azz/2$ a good Raman transfer will be achieved.

It can therefore be said that slight variations of the solution for the third degree polynomial are not critical for the protocol's performance.

\end{document}